\documentclass[12pt]{article}

\usepackage{graphics}
\usepackage{epsfig}
\usepackage{dcolumn}
\usepackage{bm}

\newcommand{\bq}{\begin{eqnarray} }
\newcommand{\eq}{\end{eqnarray} }

\begin{document}\begin{titlepage}

\begin{flushright}
\end{flushright}

\vspace*{1.0cm}

\begin{center}
{\Large {\bf X(1812) in Quarkonia-Glueball-Hybrid Mixing Scheme}}
\end{center}
\vspace*{0.3cm}
\begin{center}
 { Xiao-Gang He$^{1,2}$, Xue-Qian Li$^1$, Xiang Liu$^1$ and Xiao-Qiang Zeng$^1$}\\
\vspace*{0.3cm}
$^1$Department of Physics, Nankai University, Tianjin\\

\vspace*{0.1in}
 $^2$NCTS/TPE, Department of Physics,
 National Taiwan University,
 Taipei\\
\end{center}

 \vspace*{0.5cm}

\begin{abstract}

Recently a $J^{PC}=0^{++}$ ($X(1812)$) state with a mass near the
threshold of $\omega$ and $\phi$ has been observed by the BES
collaboration in $J/\psi \to \gamma \omega \phi$ decay. It has been
suggested that it is a $I^G = 0^+$ state. If it is true, this state
fits in a mixing scheme based on quarkonia, glueball and hybrid
(QGH) very nicely where five physical states are predicted. Together
with the known $f_0(1370)$, $f_0(1500)$, $f_0(1710)$, and
$f_0(1790)$ states, $X(1812)$ completes the five members in this
family. Using known experimental data on these particles we
determine the ranges of the mixing parameters and predict decay
properties for $X(1812)$. We also discuss some features which may be
able to distinguish between four-quark and hybrid mixing schemes.

\end{abstract}


\end{titlepage}
\newpage

\section{Introduction}

An enhancement has been observed by the BES collaboration near the
threshold of the invariant mass spectrum of $\omega\phi$ in the
radiative decay $J/\psi\rightarrow \gamma \omega\phi$. Their results
indicate the existence of a new resonant state of $J^{PC}=0^{++}$
with a mass and a width given by
$m=1812^{+19}_{-26}(\mathrm{stat})\pm 18(\mathrm{syst})
\mathrm{MeV}/\mathrm{c}^2$ and $\Gamma=105\pm 20(\mathrm{stat})\pm28
(\mathrm{syst})\mathrm{MeV}/\mathrm{c}^2$. The observed branching
ratio for $J/\psi \to \gamma \omega \phi$ is $B(J/\psi\rightarrow
\gamma X)\cdot B(X\rightarrow
\omega\phi)=(2.61\pm0.27(\mathrm{stat})\pm0.65(\mathrm{syst}))\times
10^{-4}$ \cite{bes-1812}. This resonant state is named as $X(1812)$.

Earlier the BES collaboration also reported another $J^{PC}=0^{++}$
state in the spectrum of $\pi\pi$ of $J/\psi \to \phi \pi\pi$ with a
mass of $1790^{+40}_{-30}$ MeV and a width of $270^{+80}_{-30}$ MeV,
named $f_0(1790)$. The branching ratio $B(J/\psi \to \phi f_0(1790)
\to \phi \pi\pi)$ is determined to be $(6.2\pm1.4) \times
10^{-4}$\cite{bbes}. It has been suggested that $f_0(1790)$ is a
$I^G(J^{PC}) = 0^+(0^{++})$ state. There are several other
$0^+(0^{++})$ states with mass in the range of 1 GeV to 2 GeV, these
are $f_0(1370)$ , $f_0(1500)$ and $f_0(1710)$. The states
$f_0(1370)$, $f_0(1500)$, $f_0(1710)$, $f_0(1790)$, having the same
quantum numbers with masses not far from each other, can have
significant mixing. The usual basis describing meson mixing based on
QCD picture, includes quarkonia and glueball. Without considering
excited states, the ground states of the quarkonia and glueball
basis, $N=(\bar u u+\bar d d)/\sqrt{2}$, $S=\bar s s$ and $G = gg$,
can only accommodate three $0^+(0^{++})$ states. Previous studies
have, therefore, considered three states mixing with $f_0(1370)$,
$f_0(1500)$ and $f_0(1710)$ as
members\cite{3states,shen,close-mixing,close}. The addition of
$f_0(1790)$ into the picture requires an enlargement of the basis.
In QCD, the next simplest states having the quantum numbers compared
with the quarkonia and glueball basis is the hybrid basis composed
of an anti-quark $\bar q$, a quark $q$ and a gluon $g$, i.e. $\bar q
q g$ which contains two independent $0^+$ states, $\xi_N = (\bar u
u+\bar d d)g/\sqrt{2}$ and $\xi_S = \bar s s g$. Therefore
introduction of hybrid states to accommodate $f_0(1790)$ implies the
existence of another $0^+$ state. In our recent study\cite{1790}, we
have carried out such an analysis. Since the mass of the possible
new state was not known at the time, two solutions for the
eigenstates (mainly hybrid states) were obtained with one of them
having a mass about 1760 MeV and the other about 1820 MeV. The later
case fits the new $X(1812)$ state well within the experimental
error.

We remark that the identification of $X(1812)$ as a mainly hybrid
state has an extra bonus. If $X(1812)$ is a quarkonia state,
$J/\psi\rightarrow \gamma \omega\phi$ decay is a doubly OZI
suppressed process. Thus its branching ratio should be small.  The
observed branching ratio $B(J/\psi\rightarrow \gamma X)\cdot
B(X\rightarrow
\omega\phi)=(2.61\pm0.27(\mathrm{stat})\pm0.65(\mathrm{syst}))\times
10^{-4}$ is too large to be explained. This fact indicates that
$X(1812)$ contains exotic component which allows larger branching
ratio for $J/\psi \to \gamma X(1812)\to \gamma \omega \phi$. We note
that both glueball and hybrid states can transit into a $\omega
\phi$ state without the usual OZI suppression. If indeed $X(1812)$
is mainly a hybrid state, it can naturally explain the large than
expected branching ratio for $J/\psi \to \gamma X(1812)\to \gamma
\omega \phi$. The quarkonia, glueball and hybrid (QGH) mixing scheme
proposed in Ref. \cite{1790} therefore provides a natural
description of the five members in the $0^+(0^{++})$ family
mentioned above. In this paper we study further the implications of
the QGH mixing scheme, and comment on four quark scheme for
$X(1812)$.

\section{A Scenario for QGH mixing matrix}

We now study possible structure for the QGH mixing. The effective
Hamiltonian $\mathcal{H}$ for the system cannot be calculated from
QCD yet because of complicated non-perturbative effects. There have
been some efforts to estimate the masses of hybrid mesons  by using
Constituent Gluon Model\cite{Horn}, Flux Tube Model\cite{Isgur}, Bag
Model\cite{Barnes}, QCD Sum Rules\cite{Govaerts} and also Lattice
QCD\cite{Michael}. A summary at HARDRON'95 listed the mass range for
the ground-state of hybrid as 1.3-1.8GeV\cite{KEK}. Some relevant
topics about the experimental status of hybrid states can be found
in Ref.\cite{pdg}. In Ref.\cite{Chao}, the author used the bag model
to estimate the mass ranges of scalar hybrids, and obtained
1.51-1.90 GeV for $(u\bar{u}+d\bar{d})g/\sqrt{2}$ and 2.0-2.1 GeV
for $s\bar{s}g$. Lattice calculations give $M_G$\cite{lattice} to be
in the range $1.5 \sim 1.7$ GeV. Since theoretical uncertainties on
the masses are too large to rule out a particular mass range, we
will take a more phenomenological approach assuming the QGH mixing
scheme and study some consequences of this mixing scheme. Although
it is difficult to have a precise theoretical prediction on the
mixing parameters, some simplifications can be made. One first
notices that the matrix elements $<N|\mathcal{H}|S>$ and
$<\xi_N|\mathcal{H}|\xi_S>$ are OZI suppressed and can therefore be
neglected at the lowest order approximation. The same argument
applies to $<N,S|\mathcal{H}|\xi_{N,S}>$. Possible large mixing can
occur between glueball and quarkonia, hybrid states. Since the
couplings of glueball-quarkonia, and glueball-hybrid are
flavor-independent, one has the relation $e=\langle
G|\mathcal{H}|\xi_{S}\rangle=\langle
G|\mathcal{H}|\xi_{N}\rangle/\sqrt{2}$, and $f=\langle
G|\mathcal{H}|S\rangle=\langle G|\mathcal{H}|N\rangle/\sqrt{2}$.
With the approximation described above, the mass matrix can be
expressed as
\begin{eqnarray}
M=\left (
\begin{array}{ccccc}
M_{\xi_{S}}&0&e&0&0\\
0&M_{\xi_{N}} & \sqrt{2}e & 0&0\\
e&\sqrt{2}e& M_{G}& f & \sqrt{2}f\\
0&0& f& M_{S}&0\\
0&0&\sqrt{2}f&0&M_{N}
\end{array} \right ),\label{matrix}
\end{eqnarray}
where $M_{\xi_{S}}=\langle \xi_S|\mathcal{H}|\xi_S\rangle$,
$M_{\xi_{N}}=\langle \xi_N|\mathcal{H}|\xi_N\rangle$, $M_{G}=\langle
G|\mathcal{H}|G\rangle$, $M_{S}=\langle S|\mathcal{H}|S\rangle$ and
$M_{N}=\langle N|\mathcal{H}|N\rangle$.

We parameterize the relation between the physical states and the
basis  as
\begin{eqnarray}
\left( \begin{array}{c}|X_1\rangle\\
|X_2 \rangle\\
|X_3\rangle\\
|X_4 \rangle\\
|X_5\rangle \end{array}\right)=\left( \begin{array}{c}|X(1812)\rangle \\|f_{0}(1790)\rangle\\
|f_{0}(1710)\rangle\\
|f_{0}(1500)\rangle\\
|f_{0}(1370)\rangle \end{array}\right)=U\left( \begin{array}{c} |\xi_{S}\rangle\\ |\xi_{N}\rangle\\
|G\rangle\\
|S\rangle\\
|N\rangle \end{array}\right),\;\;U=\left (
\begin{array}{ccccc}
v_{1}&w_{1}& z_{1} & y_{1}&x_{1}\\
v_{2}&w_{2}& z_{2} & y_{2}&x_{2}\\
v_{3}&w_{3}& z_{3} & y_{3}&x_{3}\\
v_{4}&w_{4}& z_{4} & y_{4}&x_{4}\\
v_{5}&w_{5}& z_{5} & y_{5}&x_{5}
\end{array} \right ).\label{unitary}
\end{eqnarray}

As $\mathcal{H}$ is not derivable and therefore neither all the
matrix elements, we need to determine them by fitting data.  The
mixing parameters $v_i$, $z_i$ and $y_i$ depend on the seven
parameters $M_{\xi_S,\xi_N,G,S,N}$, $e$ and $f$. The available data
which are directly related to these parameters are the five known
eigen-masses of $X(1812)$, $f_0(1790)$, $f_0(1710)$, $f_0(1500)$,
$f_0(1370)$. To completely fix all the parameters, more information
is needed. To this end, we use information from the ratios of the
measured branching ratios of $f_0(1790)$, $f_0(1710)$, $f_0(1500)$,
and $f_0(1370)$ to two pseudoscalar mesons listed in Table 1.
\begin{figure}[htb]
\begin{center}
\begin{tabular}{ccccc}
\scalebox{0.4}{\includegraphics{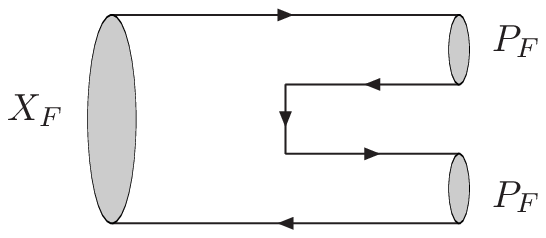}}
&\scalebox{0.4}{\includegraphics{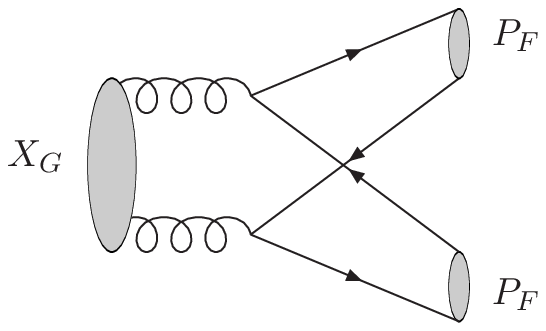}}
&\scalebox{0.4}{\includegraphics{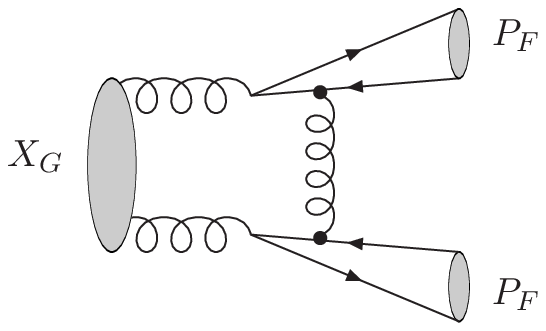}}
&\scalebox{0.4}{\includegraphics{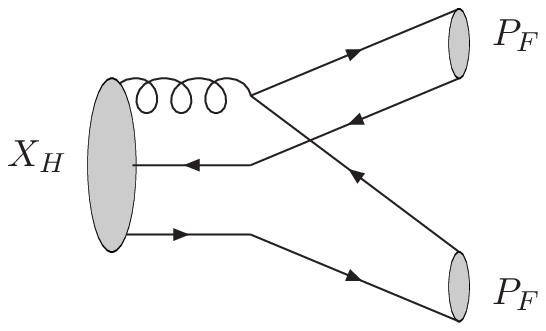}}
&\scalebox{0.4}{\includegraphics{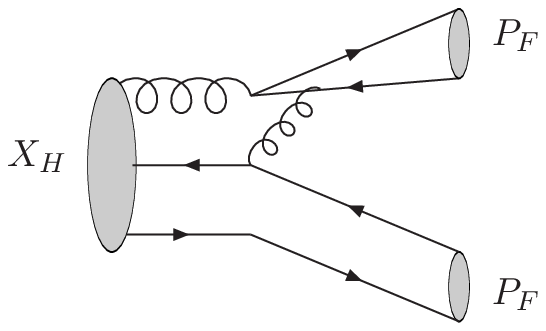}}
\\(1)&(2)&(3)&(4)&(5)\\
\scalebox{0.4}{\includegraphics{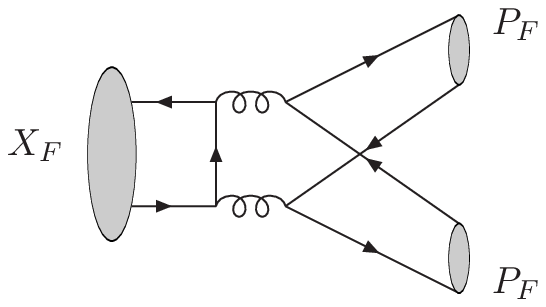}}
&\scalebox{0.4}{\includegraphics{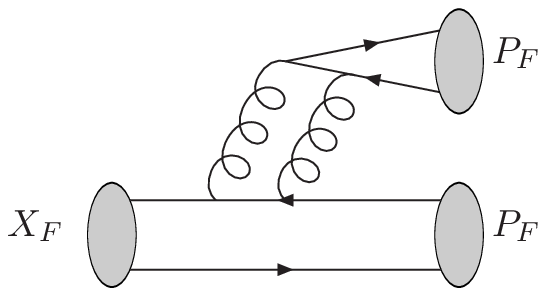}}
&\scalebox{0.4}{\includegraphics{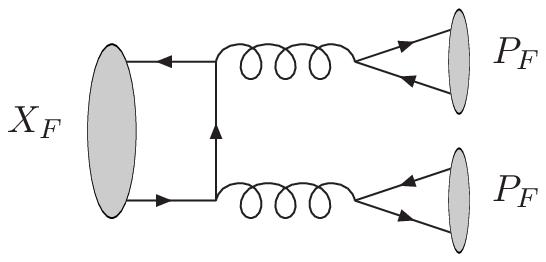}}
&\scalebox{0.4}{\includegraphics{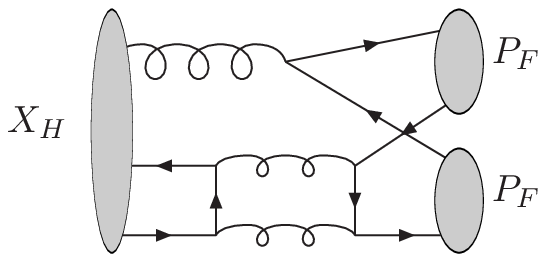}}
&\scalebox{0.4}{\includegraphics{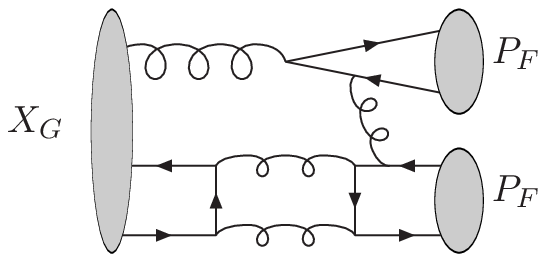}}
\\(6)&(7)&(8)&(9)&(10)\\
\end{tabular}
\end{center}
\caption{The diagrams corresponding, respectively, to terms in
eq.(\ref{decay}). The last five terms are OZI suppressed ones. The
processes of $X_i\to VV'$ can be described by the same diagrams with
the two pseudoscalar mesons in the final states replaced by two
vector mesons.} \label{diagram}
\end{figure}

The effective Hamiltonian of scalar state decaying into two
pseudoscalar mesons can be written as  \cite{lagrangian}
\begin{eqnarray}
\mathcal{H}^{PP}_{eff}&=&f_{1}\mathrm{Tr}[X_{F}P_{F}P_{F}]
+f_{2}X_{G}\mathrm{Tr}[P_{F}P_{F}]
+f_{3}X_{G}\mathrm{Tr}[P_{F}]\mathrm{Tr}[P_{F}]
+f_{4}\mathrm{Tr}[X_{H}P_{F}P_{F}]\nonumber\\
&+&
f_{5}\mathrm{Tr}[X_{H}P_{F}]\mathrm{Tr}[P_{F}]+f_{6}\mathrm{Tr}[X_{F}]\mathrm{Tr}[P_{F}P_{F}]
+ f_{7}\mathrm{Tr}[X_{F}P_{F}]\mathrm{Tr}[P_{F}]\nonumber\\
&+&f_{8}\mathrm{Tr}[X_{F}]\mathrm{Tr}[P_{F}]\mathrm{Tr}[P_{F}]
+f_{9}\mathrm{Tr}[X_{H}]\mathrm{Tr}[P_{F}P_{F}]
+f_{10}\mathrm{Tr}[X_{H}]\mathrm{Tr}[P_{F}]\mathrm{Tr}[P_{F}].\label{decay}
\end{eqnarray}
Here $X_{F,G,H}$ are the quarkonia, glueball and hybrid states.
$X_{F,H}$ are diagonal matrices $X_{F,H}=diag(X_{F,H}^1, X_{F,H}^2,
X_{F,H}^3)$. In terms of the physical component, we have
\begin{eqnarray}
&&X_F^1 = X_F^2 = {1 \over \sqrt{2}} |N\rangle = \sum_{i}{x_i\over
\sqrt{2}}X_{i},\;\;X_F^3 = |S\rangle =\sum_{i}y_{i}X_{i}, \nonumber\\
&&X_{H}^1 = X_H^2 = {1\over \sqrt{2}}|\xi_N\rangle
=\sum_{i}\frac{w_{i}}{\sqrt{2}}X_{i},\;\;X_H^3=|\xi_S\rangle = \sum_{i}v_{i}X_{i},\nonumber\\
&&X_{G}=|G\rangle = \sum_{i}z_{i}X_{i}.
\end{eqnarray}
$P_F$ is the nonet pseudoscalar mesons,
\begin{eqnarray}
P_{F}=\left(\begin{array}{ccc}
\frac{\pi^{0}}{\sqrt{2}}+\frac{x_{\eta}\eta+x_{\eta'}\eta'}{\sqrt{2}}&\pi^{+}&K^{+}\\
\pi^{-}&-\frac{\pi^{0}}{\sqrt{2}}+\frac{x_{\eta}\eta+x_{\eta'}\eta'}{\sqrt{2}}&
K^{0}\\
K^- &\bar{K}^{0}&y_{\eta}\eta+y_{\eta'}\eta'
\end{array}\right).
\end{eqnarray}
In the above,
$x_{\eta}=y_{\eta'}=(\cos\theta-\sqrt{2}\sin\theta)/\sqrt{3}$,
$x_{\eta'}=-y_{\eta}=(\sin\theta+\sqrt{2}\cos\theta)/{\sqrt{3}}$
with $\theta=-19.1^\circ$ being the $\eta-\eta'$ mixing
angle\cite{etamixing}.

The corresponding diagram representation for each term $f_i$ is
shown in Figure 1. The terms $f_{6-10}$ in the above effective
Hamiltonian describing the decay modes with two meson final states
are OZI suppressed as can be seen from Figure 1((6)-(10)). The
contributions from these terms can be neglected to a good
approximation. Within this approximation, 5 parameters (actually 4
parameters $\xi_{i}=f_{1+i}/f_1$ when considering ratios of
branching ratios) are needed to describe decay modes with two
pseudoscalar mesons in the final states.

If the $X(1812)$ state is indeed the fifth member of the QGH
mixing scheme, one has one more data point, the mass, to constrain
the parameters. Totally we now have five eigen-masses of
$f_0(1370)$, $f_0(1500)$, $f_0(1710)$, $f_0(1790)$, $X(1812)$, and
nine ratios of the branching ratios listed in Table 1\footnote{In
our fit we take the $90\%$ C.L. as 2$\sigma$ error and take the
central value to be zero for the data point for
$\Gamma(f_{0}(1710)\rightarrow
\eta\eta')/\Gamma(f_{0}(1710)\rightarrow K\bar{K})$.} to determine
the 11 parameters (7 parameters in the mass matrix plus the 4
parameters $\xi_i$ in the decay amplitudes). One therefore is able
to carry out a $\chi^2$ analysis with 3 degrees of freedom to test
the mechanism in detail.  In our fit, we also made sure that the
allowed parameter space should not result in any predicted
branching ratio to be larger than unity when data on total decay
widths of relevant particles are used.

\begin{table}[htb]
\begin{center}
\begin{tabular}{|c|c|c|} \hline
&Experiment data\cite{WA}&Best fit \\
\hline $\frac{\Gamma(f_{0}(1370)\rightarrow
\pi\pi)}{\Gamma(f_{0}(1370)\rightarrow K\bar{K})}$& $2.17\pm0.90$&
2.22\\
\hline $\frac{\Gamma(f_{0}(1370)\rightarrow
\eta\eta)}{\Gamma(f_{0}(1370)\rightarrow K\bar{K})}$& $0.35\pm0.30$&0.42\\
\hline $\frac{\Gamma(f_{0}(1500)\rightarrow
\pi\pi)}{\Gamma(f_{0}(1500)\rightarrow \eta\eta)}$&$5.56\pm0.93$& 5.45\\
\hline $\frac{\Gamma(f_{0}(1500)\rightarrow
K\bar{K})}{\Gamma(f_{0}(1500)\rightarrow \pi\pi)}$& $0.33\pm0.07$&0.32\\
\hline $\frac{\Gamma(f_{0}(1500)\rightarrow
\eta\eta')}{\Gamma(f_{0}(1500)\rightarrow \eta\eta)}$& $0.53\pm0.23$&  0.26\\
\hline $\frac{\Gamma(f_{0}(1710)\rightarrow
\pi\pi)}{\Gamma(f_{0}(1710)\rightarrow K\bar{K})}$&$0.20\pm0.03$&0.20\\
\hline $\frac{\Gamma(f_{0}(1710)\rightarrow
\eta\eta)}{\Gamma(f_{0}(1710)\rightarrow K\bar{K})}$& $0.48\pm0.19$
&0.27
\\\hline
$\frac{\Gamma(f_{0}(1710)\rightarrow
\eta\eta')}{\Gamma(f_{0}(1710)\rightarrow K\bar{K})}$&
$<0.04(90\%\;\mathrm{C.L.})$ &0.007
\\\hline
$\frac{\Gamma(f_{0}(1790)\rightarrow
\pi\pi)}{\Gamma(f_{0}(1790)\rightarrow K\bar{K})}$&$3.88^{+5.6}_{-1.9}\cite{bbes}$ &3.84\\
\hline $M_{X(1812)}$(MeV)\cite{bes-1812}&
 $1812^{+19}_{-26}(\mathrm{stat})\pm 18(\mathrm{syst})$& 1809\\\hline
$M_{f_{0}(1790)}$(MeV)\cite{bbes}&$1790^{+40}_{-30}$&1797 \\\hline
$M_{f_{0}(1710)}$(MeV)\cite{pdg}&$1714\pm5$&1714\\\hline
$M_{f_{0}(1500)}$(MeV)\cite{pdg}&$1507\pm5$&1510\\\hline
$M_{f_{0}(1370)}$(MeV)\cite{pdg}&$1350\pm 150$&1242\\\hline
\end{tabular}
\end{center}
\caption{The measured and predicted central values for branching
ratios and masses.  The minimal $\chi^2$ per degree of freedom is
1.26.} \label{fig}
\end{table}

The best fit values for relevant quantities from our $\chi^2$
analysis are listed in Tables 1 and  2. The minimal $\chi^2$ per
degree of freedom  of our fit is 1.26 indicating a good fit. The
data fitting quality has been improved compared with our previous
study. The QGH mixing scheme is a reasonable scheme to describe the
mixing of the five $0^+(0^{++})$ states. In Table 2 we also list
estimates for the 68.3\% error tolerance in the parameters by
allowing minimal $\chi^2$ per degree of freedom to float up by an
amount accordingly (with three degrees of freedom it is 1.17). We
see that the $\chi^2$ is not sensitive to $\xi_4$. More data are
need to have a better determination for these parameters.

The best fit values for the mxing matrix elements are given by
\begin{eqnarray*}
U = \left ( \begin{array}{lllll}
-0.971&-0.197&-0.106&-0.074&-0.031\\
-0.215&+0.967&+0.106&+0.081&+0.032\\
-0.087&-0.143&+0.403&+0.888&+0.146\\
+0.048&+0.070&-0.707&+0.429&-0.557\\
+0.020&+0.029&-0.562&+0.127&+0.817\end{array}\right ). \label{U}
\end{eqnarray*}

We see that the dominant component of $X(1812)$ is $s\bar{s}g$,
whereas the $(u\bar{u}+d\bar{d})g/\sqrt{2}$ is the dominant one in
$f_{0}(1790)$. The main components of $f_{0}(1710)$, $f_{0}(1500)$
and $f_{0}(1370)$ are S, glueball(G) and N, respectively.

\begin{table}[htb]
\begin{center}
\begin{tabular}{|c|c||c|c|c|c|} \hline
Parameter&Best fit and errors&Parameter&Best fit and errors
\\\hline
&&e&$20^{+8}_{-12}$  (MeV)\\
$M_{\xi_{S}}$&$1807^{+58}_{-7}$ (MeV)&f&$97^{+7}_{-6}$  (MeV)\\
$M_{\xi_{N}}$&$1794^{+7}_{-23}$  (MeV)&$\xi_{1}$&$0.83^{+0.07}_{-0.03}$ \\
$M_{G}$&$1465^{+9}_{-9}$  (MeV)&$\xi_{2}$&$0.53^{+0.28}_{-0.37}$\\
$M_{S}$&$1670^{+10}_{-11}$  (MeV)&$\xi_{3}$&$0.92^{+0.55}_{-0.73}$\\
$M_{N}$&$1336^{+17}_{-10}$  (MeV)&$\xi_{4}$& $-3.08^{+3.41}_{-1.65}$\\
\hline
\end{tabular}
\end{center}
\caption{The values for the parameters in the mass matrix $M$ and
the ratios $\xi_i = f_{1+i}/f_1$ ($i = 1 \sim 4$) in the decay
effective Hamiltonian $\mathcal{H}^{PP}_{eff}$. \label{fig-1}}
\end{table}

\section{QGH Predictions for $X(1812)$ and $f_0(1790)$ decays}

Predictions can be made for $X(1812)$ and $f_0(1790)$ decays using
the QGH mixing scheme with parameters determined in the previous
section. These predictions can be used to further test the QGH
mixing mechanism and the mixing pattern suggested. We will
concentrate on two pseudoscalar $PP'$ and two vector $VV'$ decays
here.

\noindent {\bf $X(1812) (f_{0}(1790)) \to P P'$}

The decay amplitudes for two-pseudoscalar-meson decays can be
obtained using eq.(\ref{decay}). With the numerical values
determined for the parameters we obtain
\begin{eqnarray*}
&&B(X(1812) \to \pi\pi): B(X(1812)\to K \bar K): B(X(1812) \to
\eta\eta): B(X(1812)\to \eta \eta')\nonumber\\&& = 4:37:33:0.3, \\
&&B(f_{0}(1790)\to \pi\pi):B(f_{0}(1790)\to K \bar K): B(f_{0}(1790)
\to \eta\eta): B(f_{0}(1790)\to \eta \eta')\\&& = 17:4:10:54.
\end{eqnarray*}
The above ratios also stand for $B(J/\psi \to \gamma X(1812) \to
\gamma PP')$ and $B(J/\psi \to \gamma f_0(1790)\to \gamma PP')$.

The normalization of the above branching ratios can be fixed by
using the measured value of $B(f_{0}(1710)\rightarrow
K\bar{K})=0.38^{+0.09}_{-0.19}$ \cite{pdg} and the measured widths
for the $X_i$ states. We obtain the corresponding values for
$\Gamma(X(1812)(f_{0}(1790))\to PP')$ given in Table 3. The large
branching ratios for $X(1812) \to \bar K K, \eta\eta$ and $f_0(1790)
\to \eta\eta'$ are good tests for this mechanism.

We remark that to guarantee the resultant branching ratios of $X_i
\to P P'$  to be less than unity (which must be) is a non-trivial
task since we have used experimental data for the decay widths.
The success increases our confidence on the QGH mixing scheme.

\begin{table}[htb]
\begin{center}
\begin{tabular}{|c|c|c|} \hline
$BR(X(1812)\rightarrow\pi\pi)$&$4.4\%$\\
$BR(X(1812)\rightarrow K\bar{K})$&$37.1\%$\\
$BR(X(1812)\rightarrow\eta\eta)$&$32.6\%$\\
$BR(X(1812)\rightarrow\eta\eta')$&$0.29\%$\\
$BR(f_{0}(1790)\rightarrow\pi\pi)$ &$16.8\%$\\
$BR(f_{0}(1790)\rightarrow K\bar{K})$ &$4.4\%$\\
$BR(f_{0}(1790)\rightarrow\eta\eta)$  &$9.8\%$\\
$BR(f_{0}(1790)\rightarrow\eta\eta')$ &$54.5\%$\\\hline
\end{tabular}\label{fig-2}
\end{center}
\caption{The central values for the branching ratios of $X(1812)\to
PP'$ and $f_{0}(1790)\to PP'$. }
\end{table}

\noindent {\bf $X(1812) (f_{0}(1790)) \to VV'$}

The two-vector-meson decay modes are important ones to study since
in fact the resonance $X(1812)$ is observed in the $VV'$ channel.
The effective Hamiltonian is similar to that for the Pseudoscalar
meson case with certain modifications. Corresponding to each of the
terms for $P_FP_F$ in eq.(\ref{decay}), there are two terms
$V^{\mu\nu}V_{\mu\nu}/2$ ($V^{\mu\nu} = \partial^\mu V^\nu -
\partial^\nu V^\mu$) and $V^\mu V_\mu$. Here $V$ is the nonet vector meson states,
\begin{eqnarray}
V=\left(\begin{array}{ccc}
\frac{\rho^{0}}{\sqrt{2}}+{\omega\over \sqrt{2}}&\rho^{+}&K^{*+}\\
\rho^{-}&-\frac{\rho^{0}}{\sqrt{2}}+\frac{\omega}{\sqrt{2}}&
K^{*0}\\
K^{*-} &\bar{K}^{*0}&\phi
\end{array}\right).
\end{eqnarray}

We will denote the couplings by $g_i$ and $g_ia_i$ for the two terms
respectively for $VV'$ decays, in place of $f_i$ for $PP'$ decays.
For example, the terms corresponding to
$f_{1}\mathrm{Tr}[X_{F}P_{F}P_{F}]$ will be written as
$(1/2)g_{1}\mathrm{Tr}[X_{F}V^{\mu\nu}V_{\mu\nu}]+
g_{1}a_1\mathrm{Tr}[X_{F}V^\mu V_\mu]$.
 To the leading approximation one can neglect the OZI
suppressed amplitudes $g_{6-10}$. We obtain\cite{vv-lagrangian}
\begin{eqnarray}
A(X_{i}\rightarrow \rho\rho)&=& \sqrt{3}
\bigg(\tilde g_{1}{x_{i}}+\sqrt{2}\tilde g_{2}z_{i}+ \tilde g_{4}{w_{i}}\bigg),\label{vv-1}\nonumber\\
A(X_{i}\rightarrow K^{*}\bar{K^{*}})&=& \bigg( \tilde
g_{1}{x_{i}}+\tilde g_{1}\sqrt{2}y_{i}+2\sqrt{2}\tilde g_{2}z_{i}+
\tilde g_{4}\sqrt{2}v_{i}+\tilde g_{4}{w_{i}}\bigg),\label{vv-2}\nonumber\\
A(X_{i}\rightarrow \omega\omega)&=& \bigg(\tilde g_{1}
{x_{i}}+\tilde g_{2}\sqrt{2}z_{i}+2\sqrt{2}\tilde g_{3}z_{i}+\tilde
g_{4}{w_{i}}+{2}\tilde g_{5}w_{i}
\bigg),\label{vv-3}\nonumber\\
A(X_{i}\rightarrow \omega\phi)&=& \bigg(2\sqrt{2}\tilde
g_{3}z_{i}+\sqrt{2}\tilde g_{5}v_{i}+\tilde
g_{5}{w_{i}}\bigg),\label{vv-4}
\end{eqnarray}
where $\tilde g_{j}\approx
g'_j\epsilon_{_{V_{1}}}\cdot\epsilon_{_{V_{2}}}$ with $g'_j =
g_j(p_1\cdot p_2 + a_j)$. Here we have only kept S-wave contribution
since the decays are all close to the threshold and the dominant
contribution comes from the S-wave term. With this approximation,
there is just one parameter $g'_j$ to consider for each of the
terms.

Unfortunately at present not much experimental information is
available for $VV'$ decays except $J/\psi \to \gamma X(1812)\to
\gamma \omega \phi$. Further theoretical considerations are needed
to clarify the situation and make useful predictions. To this end we
notice, from eq.(\ref{U}), that the physical state $X(1812)$ and
$f_0(1790)$ are dominated by $\xi_S = \bar s s g$ and $\xi_N = (\bar
u u + \bar d d)g/\sqrt{2}$. If the parameters $g'_{1 - 5}$ are
within a factor of o(1) order, one can neglect terms proportional to
$x_i$, $y_i$ and $z_i$ in eq.(\ref{vv-4}) for $X(1812)$ and
$f_0(1790)$ $VV'$ decays. With this approximation the decay
amplitudes depend on only two unknown parameters, $g'_4$ and $g'_5$.
The ratios $b_{1}=\Gamma(X(1812)\to \rho\rho)/\Gamma(X(1812)\to
\omega\phi)$, $b_{2}=\Gamma(X(1812)\to
\omega\omega)/\Gamma(X(1812)\to \omega\phi)$,
$b_{3}=\Gamma(X(1812)\to K^* \bar{K^*})/\Gamma(X(1812)\to
\omega\phi)$ and $b_{4}=\Gamma(f_{0}(1790)\to
\omega\omega)/\Gamma(b_{0}(1790)\to \rho\rho)$ depend on just one
parameter $\beta = g'_5/g'_4$.  In Figure 2, we show the ratios for
$b_i$ for $\beta$ varying from 0.3 to 3 for illustration. We see
that the relative branching ratios can change a large range. When
more experimental data become available, information on the
parameter $\beta$ will be extracted.

\begin{figure}[htb]
\begin{center}
\begin{tabular}{c}
\scalebox{1}{\includegraphics{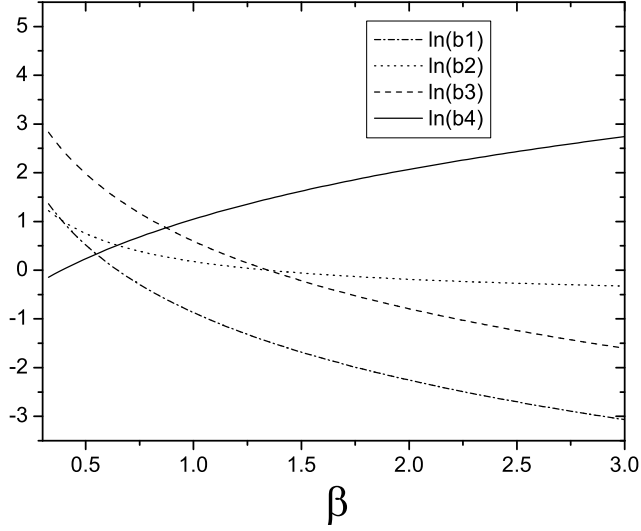}}
\end{tabular}
\end{center}
\caption{The dependence of $b_{i}$ on the parameter of $\beta$. }
\label{VV}
\end{figure}

We now make an estimate of the branching ratio for $J/\psi \to
\gamma X(1812)$. The transition matrix element of $J/\psi\rightarrow
\gamma X_{i}$ can be written as
\begin{eqnarray}
\langle\gamma X_{i}|J/\psi\rangle = x_{i}\langle\gamma
N|J/\psi\rangle+y_{i}\langle\gamma S|J/\psi\rangle+
z_{i}\langle\gamma G|J/\psi\rangle+w_{i}\langle\gamma
\xi_{N}|J/\psi\rangle+v_{i}\langle\gamma \xi_{S}|J/\psi\rangle.
\end{eqnarray}
If the SU(3) symmetry applies, we would have
\begin{eqnarray}
\langle\gamma S|J/\psi\rangle=\langle\gamma
N|J/\psi\rangle/\sqrt{2}\;\;\;\mathrm{and}\;\;\;\langle\gamma
\xi_{S}|J/\psi\rangle=\langle\gamma \xi_{N}|J/\psi\rangle/\sqrt{2},
\end{eqnarray}
and the relations \cite{gamma,close-book}  roughly hold
\begin{eqnarray}
\langle\gamma G|J/\psi\rangle:\langle\gamma
\xi_{S}|J/\psi\rangle:\langle\gamma S|J/\psi\rangle\sim
1:\sqrt{\alpha_{s}}:\alpha_{s}.\label{alphas}
\end{eqnarray}
We obtain an estimation
\begin{eqnarray}
\Gamma(J/\psi\rightarrow \gamma
X_{i})=\frac{|\mathbf{k}_{_i}|}{24\pi
M_{J/\psi}^2}\big[\alpha_{s}(\sqrt{2}x_{i}+y_{i})+\sqrt{\alpha_{s}}(v_{i}+\sqrt{2}w_{i})
+z_{i}\big]^2|M(J/\psi\rightarrow \gamma G)|^2,\label{jpsi}
\end{eqnarray}
where $\mathbf{k}_{i}$ is the three-momentum of final states in
the center of mass frame of $J/\psi$.\\

To obtain information on $|M(J/\psi \to \gamma G)|^2$ and therefore
the branching ratios for $J/\psi \to \gamma X(1812)(f_{0}(1790))$,
we use experimental data on $B(J/\psi\rightarrow \gamma
f_{0}(1710)\rightarrow \gamma K\bar{K})=8.5^{+1.2}_{-0.9}\times
10^{-4}$, $B(f_{0}(1710)\rightarrow K\bar{K})=0.38^{+0.09}_{-0.19}$
\cite{pdg}, and obtain the ranges and central values (in the
brakect) in the following with $\alpha_s =0.26$.
$$|M(J/\psi \to \gamma
G)|^2=0.005\sim 0.016 (0.007) \mathrm{GeV}^2,$$ which leads to
\begin{eqnarray}
B(J/\psi\rightarrow \gamma X(1812))&=&(0.3\sim1.0(0.4))\%,\nonumber\\
B(X(1812)\rightarrow \omega\phi)&=&(1.8\sim11.5 (6.5))\%,\nonumber\\
B(J/\psi\to\gamma f_{0}(1790))&=&(0.3\sim 0.9(0.4))\%.
\end{eqnarray}
Obviously, the numbers obtained are based on crude approximation
which should be taken as an order of magnitude estimate.

\section{Discussions and Conclusions}

In our earlier work\cite{1790}, based on the experimental
measurements on the four $0^+$ mesons ($f_0(1370), f_0(1500),
f_0(1710)$ and $f_0(1790)$), we suggested that the basis must be
enlarged to include hybrid states to have a unified description of
$0^+$ states, the QGH mixing scheme. Because there are two
independent states $(u\bar u+d\bar d)g/\sqrt 2$ and $(s\bar s)g$ for
the hybrids of isospin singlet, we predict existence of an extra
$0^+$ meson. The new state $X(1812)$ discovered recently by the BES
collaboration fits in such a picture very nicely.

Based on the ansatz for the mixing pattern of eq.(\ref{matrix}) in
the QGH scheme, we carry out a $\chi^2$ analysis to obtain the
mixing matrix and the concerned parameters in the effective
lagrangian for $0^+\rightarrow PP'$ using all avaliable experimental
data on the spectra and decay branching ratios of the five members.
We obtain a rather satisfactory result with the minimal $\chi^2$ to
be 3.79 for three degrees of freedom. This fit can explain the
relatively large branching ratio of decay mode $X(1812)\rightarrow
\phi\omega$ observed by the BES collaboration \cite{bes-1812} which
was supposed to be a double-OZI suppressed process for usual
quarkonia state.

It is noticed that after fitting the measured values of spectra and
branching ratios, we find that the masses of $M_{\xi_{S}}$ and
$M_{\xi_{N}}$ in the mixing matrix are close, because the masses of
$f_0(1790)$ and $X(1812)$ are not far apart. That is what the data
imply. The situation for regular $q-\bar q$ system is different, as
we obtain by fitting data $M_{S}-M_{N}\simeq 350$ MeV, although in
the range of the usual SU(3) breaking effect. In Ref.\cite{Chao} the
masses of the scalar hybrids in terms of the bag model as 1.51-1.90
GeV for $(u\bar{u}+d\bar{d})g/\sqrt{2}$ and 2.0-2.1 GeV for
$s\bar{s}g$. If considering the upper limit, the difference for
hybrid states is indeed very small. The closeness may be understood
that due to the gluon existence in the state, the flavor SU(3)
breaking becomes milder. Of course this allegation needs to  be
tested in the future.

With all the information available, we have made theoretical
predictions on the decay branching ratios of $f_0(1790)$, $X(1812)$
into two pseudoscalar mesons. We find that the main decay channel of
$X(1812)$ are $K\bar K$ and $\eta\eta$. If these predictions are
confirmed by experiment, it implies that the main content of
$X(1812)$ is $s\bar{s}g$. In fact, the branching ratios of other
modes are not too small and have the same order of magnitude as
$B(X(1812)\to\omega\phi)$, and can be measured in the future
experiments. Instead, among all the decay channels of
$B(f_0(1790)\to K\bar{K},\; \eta\eta)$ are relatively small, but
$B(f_0(1790)\to\eta \eta',\; \pi\pi)$ would be large.
Experimentally, the two modes $f_0(1790)\to\pi\pi$ and $f_0(1790)\to
K\bar{K}$ have been observed, so we suggest our experimental
colleagues to measure the channel $f_0(1790)\to\eta \eta',\;
\eta\eta$.

At present, very limited data about the decays of
$f_0(1370,1500,1710,1790)$ and $X(1812)$ into two vector mesons are
available, therefore we have made further approximation to estimate
related decays by keeping only the main terms $g'_4$ and $g'_5$ in
the effective lagrangian for decay amplitudes. With more data in the
future, the relevant parameters can be determined better.

We have also made a rough estimate of the branching ratios of
$J/\Psi\to\gamma X(1812)$ and $J/\Psi\to\gamma f_0(1790)$. These
results can provide useful information to our experimental
colleagues for carrying out further tests.

Before closing this section we would like to make some comments on
another alternative scenario for $X(1812)$, the four-quark state
mechanism. Four-quark state can also accommodate new $0^+(0^{++})$
particles\cite{liba}. An immediate question arises about this
scenario is that how many ground states of $0^+(0^{++})$ can be
formed with four light quarks and how to identify the dominant
component of $X(1812)$.

The number of ground states can be easily obtained by looking at the
number of isospin $I=0$ states from $\bar q_i \Gamma q_j \bar q_k
\Gamma q_l$. Here $i,j, k,l$ are color indices. $\Gamma$ indicates
combination of Dirac matrices with appropriate Lorentz indices. We
remark that when counting the number of physical $0^{+}(0^{++})$
states, the states with the same flavor structure should be counted
as one state. To find the number of $I=0$ states formed from two
quarks ($3$ of SU(3)) and two anti-quarks ($\bar 3$ of SU(3)), one
can decompose $3$ and $\bar 3$ into SU(2) isospin group to have $3 =
1+2$ and $\bar 3 = 1 + \bar 2$ and identify the $I=0$ states. There
are, naively, five possible $I =0$ states given by
\begin{eqnarray}
&&O_{\bar s s\bar s s}=\bar s s \bar s s,\;\; O_{\bar q s \bar s q}
= \bar u s \bar s u + \bar d s \bar s d,\;\; O_{\bar s q \bar q
s}= \bar s u \bar u s + \bar s d \bar d s,\nonumber\\
&& O_{(\bar q q)_0(\bar q q)_0} = (\bar u u +\bar d d)(\bar u u+\bar
d d),\;\; O_{(\bar q q)_1 (\bar q q)_1} = \bar u d \bar d u +
{1\over 2}(\bar u u - \bar d d) (\bar u u -\bar d d )+ \bar d u \bar
u d.\nonumber
\end{eqnarray}
It is clear that $O_{\bar q s \bar s q}$ is the same as $O_{\bar s q
\bar q s}$ as far as flavor contents are concerned and therefore
should be identified as the same which we will denote as $O_{\bar s
s \bar q q}$. $O_{(\bar q q)_0(\bar q q)_0}$ is the $I=0$ state
formed from two $I=0$ $\bar q q$ structures, and $O_{(\bar q q)_1
(\bar q q)_1}$ is the $I=0$ state formed from two $I = 1$ $\bar q q$
structures.

If kinematically allowed states are dominated by $O_{\bar s s\bar s
s}$, $O_{\bar s s \bar q q}$, $O_{(\bar q q)_0(\bar q q)_0}$ and
$O_{(\bar q q)_1(\bar q q)_1}$, should have their dominant decay
modes to be of the types: $(\phi \phi,\;\eta^{(')}\eta^{(')})$,
$(\phi \omega$,$\bar K^* K^* ,\;$ $\eta^{(')}\eta^{(')},\; \bar K
K)$, $(\omega\omega,\;\; \eta^{(')}\eta^{(')})$, and $(\rho \rho,
\;\pi\pi)$, respectively. The $X(1812)$ state, if dominated by a
four-quark state, should have large $O_{\bar s s \bar q q}$
component.

The above discussion shows that if the $X(1812)$ is a $0^+$
composed of four quarks, there should be another three $0^+$
states. If these states mix with quarkonia and glueball, then
there should be  seven $0^+$ states. Interesting enough, these can
accommodate $f_0 (600)$, $f_0 (980)$, $f_0 (1370)$, $f_0 (1500)$,
$f_0 (1710)$, $f_0 (1790)$ and $X(1812)$. This is different than
the QGH mixing scheme where $f_0 (600)$ and $f_0 (980)$ are left
out in the picture which may be accounted for by introducing
molecular states. The detailed mixing is difficult to study due to
lack of both experimental and theoretical information. More
theoretical and experimental studies are needed.

\vspace{1cm}

\noindent {\bf Acknowledgements}: We thank Dr. S. Jin and Dr. X.
Shen for useful discussions concerning the properties of the new
resonant state $X(1812)$.  This work is partly supported by
NNSFC and NSC.\\


\vspace{0.5cm}

\begin{thebibliography}{99}
\bibitem{bes-1812} BES Collaboration, M. Ablikim et al, arXiv: hep-ex/
{\tt0602031}.

\bibitem{bbes} BES Collaboration, M. Ablikim et al., Phys. Lett.
{\bf B607}, 243(2005).



\bibitem{3states} F. Giacosa, Th. Gutsche, V.E. Lyubovitskij and A.
Faessler, Phys. Rev. {\bf D72}, 094006 (2005); F. Giacosa, Th.
Gutsche, V.E. Lyubovitskij and A. Faessler, Phys. Lett. {\bf B622},
277-285 (2005); S. Narison, Nucl. Phys. {\bf B509}, 312-356 (1998).
\bibitem{shen} D.M. Li, H. Yu, Q.X. Shen, Commun. Theor. Phys. {\bf 34}
507-512 (2000); D.M. Li, H. Yu, Q.X. Shen, Eur. Phys. J. {\bf C19}
529-533 (2001).

\bibitem{close-mixing} F.E. Close and A. Kirk, Phys. Lett. {\bf B483}, 345-352
(2000).


\bibitem{close} C. Amsler and F.E. Close, Phys. Lett. {\bf B353},
385 (1995); C. Amsler and F.E. Close, Phys. Rev. {\bf D53}, 295
(1996).

\bibitem{1790} X.G. He, X.Q. Li, X. Liu, X.Q. Zeng, Phys. Rev. {\bf D73}, 051502 (2006),
 arXiv: hep-ph/{\tt 0602075}.

\bibitem{Horn} D. Horn and J. Mandula, Phys. Rev. {\bf D17}, 898
(1978).

\bibitem{Isgur} N. Isgur and J. Paton, Phys. Rev. {\bf D31}, 2910
(1985); N. Isgur, R. Kokoski and J. Paton, Phys. Rev. Lett. {\bf54},
869 (1985); F. E. Close and P. R. Page, Nucl. Phys. {\bf B443}, 233
(1995); T. Barnes, F.E. Close and E.S. Swanson, Phys. Rev. {\bf
D52}, 5242 (1995); F.E. Close and S. Godfrey, Phys. Lett. {\bf
B574}, 210 (2003).

\bibitem{Barnes} T. Barnes and F.E. Close, Phys. Lett. {\bf B116}, 365
(1982); M.S. Chanowitz and S.R. Sharpe, Phys. Lett. {\bf B132}, 413
(1983); M.S. Chanowitz and S.R. Sharpe, Nucl. Phys. {\bf B222}, 211
(1983).

\bibitem{Govaerts} J. Govaerts et al., Nucl. Phys. {\bf
B248}, 1 (1984); F. de Viron and J. Govaerts, Phys. Rev. Lett. {\bf
53},2207 (1984); S.L. Zhu, Phys. Rev. {\bf D60}, 014008 (1999); S.L.
Zhu, Phys. Rev. {\bf D60}, 097502 (1999).

\bibitem{Michael} C. Michael, arxiv: hep-ph/0308293; X.Q. Luo and
Y. Liu, arxiv: hep-lat/0512044; T.W. Chiu and T.H. Hsieh, arxiv:
hep-lat/0512029.

\bibitem{KEK} S. Ishida et al., KEK Preprint 95-167, NUP-A-95-15,
November 1995, {\bf H}.

\bibitem{pdg} S. Eidelman et al., Particle Data Group, Phys. Lett. {\bf B592}, 1 (2004).

\bibitem{Chao} K.T. Chao, arxiv: hep-ph/0602190.

\bibitem{lattice} N. Ishii, H. Suganuma and H. Matsufuru,
Proc. of. {\it Lepton Scattering, Hadrons and QCD},
            edited by A.W. Thomas et al.
            (World Scientific, 2001) 252;
            in {\it Lattice 2001}, Proceedings of
            the XIXth International Symposium
            on Lattice Filed Theory, Berlin, Germany,
            edited by M. M\"{u}ller-Preussker et al.
            [Nucl. Phys. B (Proc. Suppl.) {\bf 106-107}, 516
            (2002)];
            C. J. Morningstar and M. Peardon,
            Phys. Rev. D{\bf{60}}, 034509 (1999),
            and references therein; J. Sexton, A.Vaccarino and D. Weingarten,
            Phys. Rev. Lett. {\bf 75}, 4563 (1995),
            and references therein; M.J.~Teper,
            OUTP-98-88-P (1998), arXiv: hep-th/{\tt 9812187}; N. Ishii, H. Suganuma and H. Matsufuru,
Phys. Rev. {\bf D66}, 014507 (2002).

\bibitem{lagrangian} C.S. Gao, arXiv: hep-ph/{\tt
9901367}.

\bibitem{etamixing} D. Coffman et al., Phys. Rev. {\bf D38}, 2695
(1988); J. Jousset et al., Phys. Rev. {\bf D41}, 1389 (1990).

\bibitem{WA} WA102 Collaboration, D. Barberis et al., Phys. Lett.
{\bf B479}, 59 (2000).

\bibitem{vv-lagrangian} V. Cirigliano, G. Ecker, H. Neufeld and A.
Pich, JHEP {\bf 0306}, 012 (2003); S. Weinberg, Physica A {\bf 96},
327 (1979); J. Gasser and H. Leutwyler, Annals Phys. {\bf 158}, 142
(1984); J. Gasser and H. Leutwyler, Nucl. Phys. {\bf B250}, 465
(1985); G. Ecker, J. Gasser, A. Pich and E. de Rafael, Nucl. Phys.
{\bf B321}, 311 (1989); G. Ecker, J. Gasser, H. Leutwyler, A. Pich
and E. de Rafael, Phys. Lett. {\bf B223}, 425 (1989).


\bibitem{gamma}F.E. Close, G. Farrar and Z.P. Li, Phys. Rev. {\bf
D55}, 5749 (1997).

\bibitem{close-book}F.E. Close, An Introduction to Quarks and
Partons, Academic Press, London (1979).

\bibitem{liba} B. A. Lin, hep-ph/0602072.




\end{thebibliography}
\end{document}